\documentclass[12pt,epsf]{article}
\usepackage[pctex32]{graphics}
\makeatletter

\@addtoreset{equation}{section}
\newcommand{\be}{\begin{eqnarray}}
\newcommand{\ee}{\end{eqnarray}}
\newcommand{\ba}{\begin{array}}
\newcommand{\ea}{\end{array}}
\newcommand{\nn}{\nonumber}

\makeatletter \@addtoreset{equation}{section} \makeatother

\begin{document}
\vspace{1cm}
\begin{center}
~\\~\\~\\
{\bf  \LARGE Landau Free Energy and Analytic Tricritical Point in Holographic Superfluid}

\vspace{1cm}

                      Wung-Hong Huang\\
                       Department of Physics\\
                       National Cheng Kung University\\
                       Tainan, Taiwan\\
\end{center}
\vspace{1cm}
\begin{center}{\bf  \Large ABSTRACT } \end{center}
We investigate the analytical method in studying the holographic superfluid model which is described by Maxwell field minimally coupling to a charged scalar field in a fixed AdS black hole background. We  propose a method that enables us to find exact value of coefficient in the solution and thus obtain higher-order expansion of the associated Landau free energy of  the holographic superfluid with flow. We determine the critical value of superfluid velocity at the tricritical point of holographic superfluid and compare it with the numerical value. 
\vspace{3cm}
\begin{flushleft}
*E-mail:  whhwung@mail.ncku.edu.tw\\
\end{flushleft}
\newpage
\section{Introduction}
The AdS/CFT correspondence has proved to be a valuable tool for exploring strongly coupled regimes of field theories, such as in the quark-gluon plasma [1,2,3].  Recently, the focus of applying the tools of the AdS/CFT duality
has been widened to other strongly coupled systems in physics, especially to problems in condensed matter physics which was initiated in [4].  Inspired by the idea of spontaneous symmetry breaking in the presence of horizon [5], the holographic superconductors established in [6] are remarkable example where the Gauge/Gravity duality plays an important role.  

The holographic model which describes the superfluid was first proposed and studied numerically in [7].  Later, an analytic property was invested to study the model [8,9]. In a previous paper [10] we used the simple matching method [11] to analytically investigate the phase transition in  holographic superfluid.  We had shown that the first order transition can be found for some values of parameters.  However, the simple matching method does not have any variable which controls the approximation therein.  On the other hand, the method invented in [8,9] becomes exact near critical point and therefore it can exactly prove the critical exponent $\beta ={1\over 2}$ in the holographic superfluid (and  holographic superconductors). 

Note that the point, with parameter $\xi$, at which a line of first order transition becomes a line of second order transition is called as the tricritical point.  Tricritical point can be analyzed  from the Landau theory in which the free energy is written as ($M$ is the order parameter)
\be
F(T,M,\xi)=C(T,\xi)+\mu (T,\xi) M^2+V(T,\xi) M^4+ \kappa (T,\xi) M^6+...
\ee
where $\kappa (T,\xi) > 0$ to ensure stability. The second-order phase transition is found if $\mu (T,\xi)<0$ while $V(T,\xi)  >0$.   The first-order phase transition is found if $\mu (T,\lambda)>0$ while $V(T,\xi)<0$.    At tricritical point both of  $\mu (T,\xi)$ and $V(T,\xi)$ become zero [12].  

In this paper we will extend the analytic results in [8] to determine the tricritical point in  holographic superfluid where the parameter $\xi$ is the flow velocity.   In section II, we first slightly detail the analytic method of [8]. We see that, for example, when we have already found function $At_{20}(u)$ in (2.14) then, in convention, we need to know the exact solution of $\rho_{30}(u)$ to determine the coefficient $c_1$ in (2.14). In contrast to the standard method, we propose a simple trick which enables us to find the exact value of coefficient $c_1$  without knowing the exact solution of $\rho_{30}(u)$ which couples to it.  Our trick is described in  the formulas (2.45) and (2.46). 

  Using the crucial trick we are able to find the higher-order expansion solution in the holographic superfluid with flow in section III.  In section IV we calculate the associated Landau free energy and find that the critical value of superfluid velocity at the tricritical point of holographic superfluid is $\xi=0.887$ which is close to the numerical value $\xi=0.653$ [7].   A short conclusion is given in the last section.
\section {Analytic Method  in Holographic Superfluid}
The holographic superconductor studied in [7] is describe by a Maxwell field $A_\mu$ minimally coupling to a charged scalar field $\Psi$.  The action is
\be
S=\int d^5x \sqrt{-g}\Big( -{1\over 4 e^2} F_{\mu\nu}F^{\mu\nu}-|\partial_\mu \Psi- i A_\mu|^2-m^2|\Psi|^2 \Big)
\ee
The fields are propagating on a fixed AdS black hole background
\be
ds^2 = {1\over u^2}\Big[- f(u)dt^2 +dx^2+dy^2+dz^2 +{du^2\over f(u)}\Big],~~~f(u)=1-{u^4\over u_h^4}
\ee 
The associated field equations are
\be
{1\over \sqrt{-g}}\Big(\partial_\mu-iA_\mu\Big) \sqrt{-g}\Big(\partial^\mu\Psi-iA^\mu\Psi\Big)-m^2\Psi&=&0\\
{1\over \sqrt{-g}}\partial_\mu\Big(\sqrt{-g}~F^{\mu\nu}\Big) -ie^2g^{\nu\lambda}\Big(\Psi^*(\partial_\lambda-i A_\lambda)\Psi-\Psi(\partial_\lambda+i A_\lambda)\Psi^*\Big)&=&0
\ee
After expressing  the bulk scalar as $\Psi={1\over \sqrt 2}\rho e^{i\varphi}$ and making a gauge transformation $A_\mu\rightarrow A_\mu+\partial_\mu\varphi$ the phase $\varphi$ will disappears from field equations and u-component of field equation give trivial solution of $A_u=0$ [7].  The remained field equations are 
\be
0&=&u^3\partial_u\Big[{f\over u^3}\rho'\Big]-\Big(A_i^2-{A_t^2\over f}+{m^2\over u^2}\Big)\rho \\
0&=&u\partial_u\Big[{A_t'\over u}\Big]-{\rho^2\over u^2 f}A_t\\
0&=&u\partial_u\Big[{f A_i'\over u}\Big]-{\rho^2\over u^2 }A_i
\ee
which we call as $\rho$ equation, $A_t$ equation and $A_i$ equation respectively. 

From the usual rules of the AdS/CFT duality [1,2,3] the boundary value of  bulk gauge field $A_\mu$ acts as a source for a conserved current $J_\mu$ and the
near boundary data of the scalar field $\Psi$ sources a scalar operator.  Also, the boundary value of $J_0$ is the chemical potential and  $J_i$ is the superfluid velocity. 

Hereafter we will use the unit $e=1$ and $m^2=-4$ which yields an operator of dimension two. The horizon is at $u_h=1$ and boundary is at $u=0$. Using the rotational invariance we can take $A_i = (A_x, 0, 0)$ without loss of generality. 
\subsection{Exact-Solution Method}
We first slightly detail the analytic method of [8] and then use it to illustrate a simple trick which enables us to find some exact parameters in the model. We begin with the case of $A_x=0$. 

Step 1. First, let $\rho=0$ in  $A_t$ equation we can find  the exact solution
\be
A_t(u)=c_1+c_2 u^2
\ee
Under the property that function $A_t(u)$ must zero on horizon, i.e. $u=1$, we find that $A_t(u)=c_1(1- u^2)$. 

Step 2.  After substituting above solution into $\rho$ equation we can find the following exact solution [8]
\be
A_t(u)&=&2(1- u^2)\\
\rho(u)&=&=\frac{c_1 u^2}{u^2+1}+\frac{c_2 u^2 \left(\log (u)-\log \left(1-u^2\right)\right)}{u^2+1}
\ee
As function $\rho(u)$ shall be regular on horizon we need to let $c_2=0$. Thus we get the regular solution (set $c_1 \equiv \epsilon$)
\be
\rho(u)=\frac{\epsilon~ u^2}{u^2+1}
\ee
in which $\epsilon$ is an integration constant used to describe the order parameter. 

  We will now find the analytical series solutions of $A_t(u)$ and $\rho(u)$ expanded in the parameter $\epsilon$ :
\be
A_t(u)&=&At_{00}(u)+\epsilon^2At_{20}(u)+\epsilon^4At_{40}(u)+...\\
\rho(u)&=&\epsilon~\rho_{10}(u)+\epsilon^3\rho_{30}(u)+\epsilon^5\rho_{50}(u)+...
\ee
The strategy invented in [8] is that in solving  the differential equations we need to require the boundary condition that the $O(u^2)$ term in $\rho i0(u)$ vanish (for $i > 1$) while $Ai0(u)$ is allowed to be nonzero. After having found the lowest-order solutions $At_{00}(u)=2(1-u^2)$ and $\rho_{10}(u)=\frac{u^2}{u^2+1}$ we turn to find higher-order solutions.

Step 3. Substituting $\rho(u)=\epsilon~\rho_{10}$ into $A_t$ equation we can find that 
\be
\epsilon ^2At_{20}(u) =\frac{1}{2} c_1 \left(u^2+1\right)+\frac{1}{8} \left(-8 c_1-\epsilon ^2\right)+\frac{\epsilon^2}{4 u^2+4}
\ee
after imposing the property that  function $At_{20}(u)$ must zero on horizon.  However, we could not determine the value of  $c_1$  by the $At_{20}(u)$ equation itself.  In fact it shall be determined by higher-order $\rho$ equation in the following step.  (This is because that $At_{20}(u)$ couples to $\rho_{30}(u)$.)

Step 4.  After substituting the above solution of $At_{20}(u)$ (with undetermined constant $c_1$) into $\rho$ equation we can find the exact solution
\be
\epsilon ^3\rho_{30}(u) &=& \frac{u^2}{96 \Big(u^2+1\Big)^2}
 \Big[4\epsilon ^3+4 \Big(u^2+1\Big) \Big(24 c_3+\epsilon ^3\Big) \log
   (u)\nn\\
&&+\Big(u^2+1\Big) \Big(-24 c_1 \epsilon -96 c_3+\epsilon ^3\Big) \log
   \Big(1-u^2\Big)\nn\\
&&+24 c_1 u^2 \epsilon  \log \Big(u^2+1\Big)+24 c_1 \epsilon  \log
   \Big(u^2+1\Big)\nn\\
&&+96 c_2 u^2+96 c_2-3 u^2 \epsilon ^3 \log \Big(u^2+1\Big)-3 \epsilon ^3\log \Big(u^2+1\Big)\Big]
\ee
in which $c_2$ and $c_3$ are the new integration constants to be determined. 

Imposing the regularity on boundary and on horizon we find that 
\be
c_1&=& \frac{5 \epsilon^2}{24}\\
c_3&=& -\frac{\epsilon^3}{24}
\ee
Using above value of $c_1$ we find that [8]
\be
At_{20}(u) =\frac{5 u^4-6 u^2+1}{48 \left(u^2+1\right)}
\ee
Now, we see that $A_t=At_{00}+\epsilon^2~At_{20}= (\mu_c+\epsilon^2~\delta \mu) +{\cal O}(u^2)\equiv \mu +{\cal O}(u^2)$, with $\mu_c=2$ and $\delta \mu={1\over 48}$.  Thus, $\epsilon\approx \sqrt{\mu-\mu_c\over \delta\mu}$ which implies $\epsilon\approx \sqrt{T_c-T}$ after restoring dimension by $\mu\rightarrow \mu/\pi T$.  This proves that the critical exponent $\beta ={1\over 2}$ [8].\\

To proceed, the boundary condition that the $O(u^2)$ term in $\rho_{30}(u)$ vanishes could determine $c_2$ and we find that [8]
\be
\rho_{30}(u) =\frac{u^2 \left[\left(u^2+1\right) \log \left(u^2+1\right)-2 u^2\right]}{48\left(u^2+1\right)^2}
\ee
 
Step 5. In the same way, substituting above solutions into $A_t$ equation we can find that 
\be
\epsilon ^4At_{40}(u) ={1\over1152}\Big[576 c_2 \Big(u^2-1\Big)+\epsilon ^4 \Big(\frac{7 u^2\Big(u^2-1\Big)}{\Big(u^2+1\Big)^2}-6 \log (2)\Big)+\frac{12 \epsilon ^4 \log\Big(u^2+1\Big)}{u^2+1}\Big]\nn\\
\ee
after imposing the property that  function $At_{40}(u)$ must zero on horizon. 

Step 6.  As before, the value of  $c_2$ in $At_{40}(u)$ shall be determined by higher-order $\rho_{50}(u)$ equation. After solving the equation exactly and imposing the proper boundary we can find that
\be
c_2=\frac{253 \epsilon ^4-624 \epsilon ^4 \log (2)}{27648}
\ee
and thus we can get the following higher-order solutions: 
\be
At_{40}(u)&=&{1\over55296}\Big[\Big(u^2-1\Big) [253-624 \log (2)]+48 \Big[\frac{7 u^2 (u^2-1)}{(u^2+1)^2}-6 \log (2)\Big]\nn\\
&&+\frac{576 \log(u^2+1)}{u^2+1}\Big]\\
\rho_{50}(u)&=&{u^2 \over 55296 \Big(u^2+1\Big)^3}
\Big[96 \Big(u^2+1\Big)^2   {Li}_2\Big(\Big(-\frac{1}{2}-\frac{i}{2}\Big) (u-1)\Big)\nn\\
&&+96 \Big(u^2+1\Big)^2{Li}_2\Big(\Big(-\frac{1}{2}+\frac{i}{2}\Big) (u-1)\Big)-216 \Big(u^2+1\Big)^2{Li}_2\Big(-u^2\Big)\nn\\
&&+96 \Big(u^2+1\Big)^2{Li}_2\Big(\Big(\frac{1}{2}-\frac{i}{2}\Big) (u+1)\Big)+96 \Big(u^2+1\Big)^2 {Li}_2\Big(\Big(\frac{1}{2}+\frac{i}{2}\Big) (u+1)\Big)\nn\\
&&-20 \pi ^2 u^4-64 u^4-40\pi ^2 u^2-169 u^2+4 \Big(-15 \Big(u^2+1\Big)^2 \log ^2\Big(u^2+1\Big)\nn\\
&&+2\Big(\Big(7 u^2+32\Big) u^2-12 i \pi  \Big(u^2+1\Big)^2-12 \Big(u^2+2\Big) u^2\log (u+1)\nn\\
&&-12 \Big(u^2+1\Big)^2 \log (1-u)+54 \Big(u^2+1\Big)^2 \log (u)-30
   \Big(u^2+1\Big)^2 \log (2)\nn\\
&&-12 \log (u+1)\Big) \log \Big(u^2+1\Big)+24\Big(\Big(u^2+2\Big) u^2 \Big(\log \Big(\Big(\frac{1}{2}+\frac{i}{2}\Big) (1+iu)\Big)\nn\\
&&+\log ((-1-i) (u+i))\Big) \log (u+1)\nn\\
&&+\Big(u^2+1\Big)^2 (\log (1-u)+i \pi)
   \Big(\log ((1+i) (u-i))\nn\\
&&+\log \Big(\Big(\frac{1}{2}-\frac{i}{2}\Big)(u+i)\Big)\Big)\Big)+24 \log \Big(\Big(\frac{1}{2}+\frac{i}{2}\Big) (1+i u)\Big)\log (u+1)\Big)\nn\\
&&+48 \Big(u^2+1\Big)^2 \log ^2(2)-432 \Big(u^2+1\Big)^2 \log (1-i u)
   \log (u)\nn\\
&&-432 \Big(u^2+1\Big)^2 \log (1+i u) \log (u)+200 \log \Big(u^2+1\Big)\nn\\
&&+96 \log   ((-1-i) (u+i)) \log (u+1)-20 \pi ^2\Big]
\ee
in which ${Li}_2(x)$ is the  polylogarithm function.  All above exact solutions, which had presented in [8], are necessary to find the solution in next section.

Note that the  coefficient $c_1$  of  $At_{20}(u)$ in step 3 is determined in step 4, which is based on the existence of the exact solution of $\rho_{30}(u)$ equation. In the same way,  coefficient $c_2$  of  $At_{40}(u)$ in step 5 is determined in step 6 by the existence of the exact solution of $\rho_{50}(u)$ equation.  However, sometimes (in the case with superfluid flow considered in next section) we could not find the exact solution of higher-order $\rho$ solution while we need to determine the relevant coefficient, $c_1$ in lower-order $At(u)$. In this case we can use the following trick.
\subsection{Simple-Trick Method}
In this subsection we will show how one can get the constant $c_1$ in $At_{20}(u)$ solution without knowing the exact solution of $\rho_{30}(u)$. 

In general the differential equation associated with $\rho_{30}(u)$ and $\rho_{50}(u)$, or $\rho_{52}(u)$ in the next section can be expressed as 
\be
\rho_{30}''(u) +K(u)\rho_{30}' (u)+L(u)\rho_{30}(u)=G(u,c_1)
\ee
where $G(u,c_1)$ contains coefficient $c_1$ of  $At_{20}(u)$  (or other parameters $c_i$) which is to be determined. 

We can solve above equation by the standard method of variation of parameters. As associated homogeneous differential equation has two exact solutions
\be
y_1(u)&=&\frac{u^2 \log (u)}{u^2+1}\\
y_2(u)&=&\frac{u^2 \left[\log (u)-\log \left(1-u^2\right)\right]}{u^2+1}
\ee
the general solution is
\be
\rho_{30}(u)=\alpha~y_1(u)+\beta~y_2(u) + A(u) y_1(u)+B(u) y_2(u)
\ee
in which $\alpha$ and $\beta$ are arbitrary constants and 
\be
A(u)&=&\int du{- y_2(u) G(u,c_1)\over W(y_1(u),y_2(u))}\\
B(u)&=&\int du{ y_1(u) G(u,c_1)\over W(y_1(u),y_2(u))}
\ee
where $W(y_1(u),y_2(u))$ is the Wronskian of $y_1(u),y_2(u)$ defined by
\be
W(y_1(u),y_2(u))&=&\left| {\begin{array}{cc}
y_1(u)& y_2(u) \\
y_1'(u)& y_2'(u)\\
  \end{array} } \right|
\ee
Note that we can ignore the  integration constants in above integrations since including them would merely regenerate terms already present in the homogeneous  solution.

While in this step we have exact function form of $G(u,c_1)$ the function itself  may be too complex to be integrated exactly and we could not determine $c_1$ therefore. In this case we can use the following trick.

First, we know that to determine the coefficient $c_1$, as that do in step 4, we need to use the two boundary conditions: (1)  $O(u^2)$ term in $\rho_{30}$ vanishes. (2) $\rho_{30}(u)$ is regular on horizon and boundary. 

Next, as these properties are these near $u=0$ and $u=1$ one intents to guess   that it merely needs to find the functions of A(u) and B(u) near $u=0$ and $u=1$ to determine the coefficient $c_1$.  For example, in above case we can expand the function of $G(u,c_1)$ about u=0 and then preform the integration to get 
\be
A^{series}(u)&\stackrel{u\approx 0}{=}&-u \left(\frac{1}{2} \epsilon ^3 \log (u)-2 c_1~\epsilon \log (u)\right)+{\cal O}(u^3)\\
B^{series}(u)&\stackrel{u\approx 0}{=}&u\left(\frac{\epsilon ^3}{2}-2 c_1~\epsilon \right)+{\cal O}(u^3)
\ee
which can be called as series-expansion solution.  Thus
\be
\rho_{30}^{series}(u)\stackrel{u\approx 0}{=}u^2 \left(\alpha+\beta \log (u)\right)+{\cal O}\left(u^4\right)
\ee
and two boundary  conditions lead to $\alpha=\beta=0$. 

In the same way, near u=1 it gives solution
\be
A^{series}(u)&\stackrel{u\approx 1}{=}&\frac{1}{64} (1-u)^2 \epsilon[2 \log (2(1-u))-1) \Big(\epsilon ^2-8 c_1\Big)\Big]+{\cal O}((1-u)^3)\\
B^{series}(u)&\stackrel{u\approx 1}{=}&\frac{1}{32} (1-u)^2 \epsilon  \left(\epsilon ^2-8 c_1\right)+{\cal O}((1-u)^3)
\ee
Thus
\be
\rho_{30}^{series}(u)&\stackrel{u\approx 1}{=}&\frac{1}{2} \Big[-\beta \log (1-u)+\alpha-\beta \log (2)\Big)+\frac{1}{16} (1-u) \Big( \beta(8 \log (1-u)-4 \nn\\
&&+\log (256)-8\alpha\Big]+{\cal O}((1-u)^2)
\ee
and two boundary conditions lead to $\beta=0$.  However, we could not determine the value of coefficient $c_1$ in $At_{20}(u)$.  Let us explain the reason in below.

Consider the exact integration of  $A(u)$ and $B(u)$ 
\be
A^{exact}(u)&=&{\epsilon \over 96\Big(u^2+1\Big)^3}\Big(\epsilon ^2 \Big(\Big(u^2-1\Big)^2 \Big(u^2+5\Big) \log \Big(1-u^2\Big)\nn\\
&&+\Big(u^2+1\Big)^2 \Big(4-3 \Big(u^2+1\Big) \log
   \Big(u^2+1\Big)\Big)+4 \Big(u^4+3 u^2+6\Big) u^2 \log (u)\Big)\nn\\
&&-24  \Big(u^2+1\Big) c_1 \Big(4 u^2 \log (u)+\Big(u^2-1\Big)^2 \log
   \Big(1-u^2\Big)\nn\\
&&-\Big(u^2+1\Big)^2 \log \Big(u^2+1\Big)\Big)\Big)\\
B^{exact}(u)&=&\frac{\epsilon  \left(\left(3 u^2-1\right) \epsilon ^2-24 \left(u^4+u^2\right) c_1\right)}{24\left(u^2+1\right)^3}
\ee
Expanding near u=0 and u=1 give
\be
A^{exact}(0)&=&-\frac{\epsilon ^3}{24}\\
A^{exact}(1)&=&-\frac{1}{96} \left(24 \epsilon  \log (2) c_1+2 \epsilon ^3-3 \epsilon ^3 \log(2)\right)\\
B^{exact}(0)&=&-\frac{\epsilon ^3}{24}\\
B^{exact}(1)&=&\frac{1}{192} \epsilon  \left(2 \epsilon ^2-48 c_1\right)
\ee
We see that $A^{exact}(0)\ne A^{exact}(1)$ and $B^{exact}(0)\ne B^{exact}(1)$ while the series-expansion solutions presented in above give $A^{series}(0)=A^{series}(1)=B^{series}(0)=B^{series}(1)=0$.  Let us clarify the subtle now.

Note that, as have mentioned before, we have ignored the integration constants   in integrating to get functions $A(u)$ and $B(u)$, since including them would merely regenerate terms already present in the homogeneous general solution.  Thus we can arbitrary add $A^{exact}(u)$ and $B^{exact}(u)$ any constant $C_A$ and $C_B$ and this surely would not change the solution property. In fact, these constants will  be shifted to the constants in the coefficients  $\alpha$ and $\beta$ in above equations. 

 Now, we need to notice that the differences between $A^{exact}(0)$ and $A^{exact}(1)$ and between $B^{exact}(0)$ and $B^{exact}(1)$ are already fixed by the exact solution and it is irrelevant to the constant $C_A$ or $C_B$.

Thus, the inconsistent  properties that the series-expansion solutions give  $A^{series}(0)=A^{series}(1)=B^{series}(0)=B^{series}(1)=0$ is the consequence that the integration constants adopted in the case of near $u=0$ and $u=1$ are different.  This is wrong as they shall be already fixed by the exact solution. 

Due to the difference between the series expansions about u=0 and u=1 we have to correct it by compensating the difference between them and add the following values
\be
A_c&=&\int_0^1 du{ -y_2(u) G(u,c_1)\over W(y_1(u),y_2(u))}\\
B_c&=&\int_0^1 du{ y_1(u) G(u,c_1)\over W(y_1(u),y_2(u))}
\ee
into the series-expansion solution and corrected forms of series-expansion solutions near u=0 and u=1 are
\be
\rho_{30}(u)&\stackrel{u\approx 0}{=}&\alpha~y_1(u)+\beta~y_2(u) + (A^{series}(u)+A_c) ~y_1(u)+(B^{series}(u)+B_c)~ y_2(u)\nn\\
\\
\rho_{30}(u)&\stackrel{u\approx 1}{=}&\alpha~y_1(u)+\beta~y_2(u) + A^{series}(u) ~y_1(u)+B^{series}(u) ~y_2(u)
\ee
Note that above trick can be applied to arbitrary differential equation  in any model.

As examples we will apply above  method to the following two cases. Note that in our model, dues to the special function forms of $y_1(u)$ and $y_2(u)$, the functions $A^{series}$ and $B^{series}$ can be neglected in the leading-order expansion and thus equations (2.48), (2.50) and (3.13) have a same form. 
\subsection{Examples}
In the first example, assuming that we have arrived the step 3 and found $At_{20}(u)$ which has an undetermined parameter $c_1$.  After using the above formula we find that 
\be
\rho_{30}(u)&\stackrel{u\approx 0}{=}&\frac{1}{96} u^2 \Big[(96 \beta +5\epsilon ^3-24 \epsilon  c_1)\log(u)+...\Big]+{\cal O}(u^4)\\
\rho_{30}(u)&\stackrel{u\approx 1}{=}&\frac{1}{2} \Big[\alpha -\beta \log(2(1-u))\Big]+{\cal O}((1-u))
\ee
Then, the two boundary conditions lead to $\beta =0$ and $c_1={5\epsilon^2\over 24 }$ which is consistent with that calculated from exact integration in step 4.

In the second example, assuming that we have arrived the step 5 and found $At_{40}(u)$ which has an  undetermined parameter $c_2$.  After using the above formula we  find that 
\be
\rho_{50}(u)&\stackrel{u\approx 0}{=}&[-55296 \epsilon  c_2+221184 \beta+506 \epsilon ^5-1248 \epsilon ^5 \log (2)]\log(u)+{\cal O}(u^4)\\
\rho_{50}(u)&\stackrel{u\approx 1}{=}&\frac{1}{2} \Big[\alpha -\beta \log(2(1-u))\Big]+{\cal O}((1-u))
\ee
Then, the two boundary conditions lead to $\beta =0$ and $c_2=(253 \epsilon ^4-624 \epsilon ^4 \log (2))/27648$ which is consistent with that calculated from exact integration in step 6.
\\

Note that, even if  the function $G(u,c_1)$ was too complex to be integrated exactly the numerical integration from $u=0$ to $u=1$ can be calculated easily and thus get $A_c$ and $B_c$. Also, after  expanding the function of $G(u,c_1)$ about u=0 or u=1 we can easily preform the integration to get $A^{series}(u)$ and $B^{series}(u)$. Thus, above formula can be easily applied to any cases and we will  use it in next section.
\section {Analytic Method  in Holographic Superfluid with Flow}
We now consider the holographic superfluid with flow velocity $\xi$ which is treated as an another small parameter. 
\subsection{Exact-Solution Method}
The general solutions can be expanded as 
\be
A_t(u)&=&\Big[At_{00}(u)+\epsilon^2At_{20}(u)+\epsilon^4At_{40}(u)\Big]\nn\\
&&+\Big[At_{02}(u)+\epsilon^2At_{22}(u)+\epsilon^4At_{42}(u)\Big]\xi^2+...
\\
A_x(u)&=&\Big[Ax_{01}(u)+\epsilon^2Ax_{21}(u)+\epsilon^4Ax_{41}(u)\Big]\xi+...\\
\rho(u)&=&\Big[\epsilon~\rho_{10}(u)+\epsilon^3\rho_{30}(u)+\epsilon^5\rho_{50}(u)\Big]\nn\\
&&+\Big[\epsilon~\rho_{12}(u)+\epsilon^3\rho_{32}(u)+\epsilon^5\rho_{52}(u)\Big]\xi^2+...
\ee
Using the exact-solution method we can find 
\be
\rho_{12}(u)&=&-\frac{ u^2   \log \left(u^2+1\right)}{4 \left(u^2+1\right)}\\
\rho_{32}(u)&=&{u^2 \over 2304 \Big(u^2+1\Big)^2}\Big[-72 \Big(u^2+1\Big)
  {Li}_2\Big(\Big(-\frac{1}{2}-\frac{i}{2}\Big) (u-1)\Big)\nn\\
&&-72 \Big(u^2+1\Big){Li}_2\Big(\Big(-\frac{1}{2}+\frac{i}{2}\Big) (u-1)\Big)+120 \Big(u^2+1\Big){Li}_2\Big(-u^2\Big)\nn\\
&&-72 \Big(u^2+1\Big)
   {Li}_2\Big(\Big(\frac{1}{2}-\frac{i}{2}\Big) (u+1)\Big)-72 \Big(u^2+1\Big)
  {Li}_2\Big(\Big(\frac{1}{2}+\frac{i}{2}\Big) (u+1)\Big)\nn\\
&&+104 u^2+3\Big(u^2+1\Big) \Big(5 \pi ^2-12 \log ^2(2)\Big)+4 \log \Big(u^2+1\Big)\Big(-u^2\nn\\
&&+6 \Big(u^2+1\Big) \log \Big(u^2+1\Big)+36 \Big(u^2+1\Big) \log(2)-19\Big)\Big]\\
Ax_{01}(u)&=&1\\
Ax_{21}(u)&=&-\frac{ u^2 }{8 u^2+8}\\
Ax_{41}(u)&=&{1 \over 4608}\Big[-\pi ^2+12 {Li}_2\Big(\frac{1}{2} \Big(u^2+1\Big)\Big)+6 \Big[\frac{\Big(2 u^2-3\Big) u^2}{\Big(u^2+1\Big)^2}+\log ^2(2)\Big]+6  \Big[\frac{4}{u^2+1}\nn\\
&&-2 \log \Big(1-u^2\Big)\Big({\log2\over \log (1+u^2)}-1\Big)-\log\Big(u^2+1\Big)\Big] \log \Big(u^2+1\Big)\Big]\\
At_{02}(u)&=&\frac{1}{4} \left(2-2 u^2\right)\\
At_{22}(u)&=&{1\over 288 \Big(u^2+1\Big)}\Big[\Big(u^4 (45 \log (2)-28)+18 u^2 (1+\log (2))\nn\\
&&-36 \log\Big(u^2+1\Big)+10-27 \log (2)\Big)\Big]\\
At_{42}(u)&=&{1\over 110592}\Big(55296 \Big(u^2-1\Big) C_2+\xi ^2 \epsilon ^4 \Big(6224+2880u^2+\frac{280}{(u-i)^2}\nn\\
&&-(3228+648 i) \log (-1+i)\Big)+......({\rm many~terms})
\ee
in which $At_{42}(u)$ function is too lengthy to be written completely in here.  Note that the functions $At_{22}(u),~At_{42}(u),~Ax_{41}(u)$ and $\rho_{32}(u)$, which is necessary to find the tricritical point, did not calculate in [8].

  Note that we have also used the simple-trick method to determine  $At_{22}(u)$ and find that it is consistent with the exact-solution method.
 
\subsection{Simple-Trick Method}
To proceed we need to determine the coefficients $C_2$ in $At_{42}(u)$ by solving $\rho_{52}(u)$ equation exactly which, however, is an impossible task because the function form of $At_{42}(u)$ are very lengthy. Thus, we turn to use the simple-trick method. 

We then find that 
\be
\rho_{52}(u)&\stackrel{u\approx 0}{=}&u^2 \Big[A_c + \alpha + B_c \log(u) + \beta \log(u)\Big]+{\cal O}(u^4)\\
\rho_{52}(u)&\stackrel{u\approx 1}{=}&\frac{1}{2} \Big[\alpha -\beta \log(2(1-u))..\Big]+{\cal O}((1-u))
\ee
where
\be
A_c &=&-0.005336... +i\times 0.001063... +C_2\times 0.17328...
\\
B_c &=& 0.0050231... + i\times 0.0015339... -  C_2\times 0.25
\ee
in which $i\equiv \sqrt{-1}$.  The constants in above are numerical exact values, for example
\be
&& \int_0^1 du~\frac{2 u \left(u^2-1\right) \left(\log (u)-\log
   \left(1-u^2\right)\right)}{\left(u^2+1\right)^3}={1\over4} \log(2)=0.17328...\nn\\
&&\int_0^1 du~\frac{2 u \left(u^2-1\right) }{\left(u^2+1\right)^3}=-{1\over4}
=0.25
\ee
Others are coming from the integration of lengthy terms in $At_{42}(u)$, which are too cumbersome to be written down in here.

Then, the two boundary conditions lead to 
\be  \alpha &=&- A_c\\
        \beta &=&- B_c =0
\ee  
which gives
\be
\alpha &=& -0.00185521...\\
C_2&=&  0.0200924... + i\times 0.00613592...
\ee
and we finally get 
\be
At_{42}(u)&=&0.00244113... - u^2\times 0.00170545...+{\cal O}(u^4)
\ee
It is interesting to see that the primitive function form of $At_{42}(u)$ obtained in the exact solution is a complex function.  On other hand, the coefficient $C_2 $, within $At_{42}(u)$, found in simple-trick method is a complex number.  It is interesting to see that, after substituting complex number  $C_2 $ into  complex function $At_{42}(u)$ the final function form of $At_{42}(u)$ becomes real. (In fact, we get an imaginary part in $At_{42}(u)$, which is  of order ${\cal O}(10^{-18})$ and is neglected therefore.) Note that in above we only present the leading expansion as it is sufficient to calculate the free energy to second order of flow velocity, as explained in below.
\section{Landau Free Energy and Critical Velocity}
The free energy of the field theory can be evaluated  by the on-shell value of the action (2.1).  The final form is [8] 
\be
S_{\rm on-shell}=\int d^4x\Big[{g^{uu}\sqrt {-g}\over 2}(g^{\mu\nu}A_\mu A'_{\nu}+\rho\rho')|_{u=\delta}+{1\over 2}\int_\delta^{u_h}du~\sqrt{-g} g^{\mu\nu}A_\mu A_\nu\rho^2\Big]
\ee
The first part need not integration while second part need to do integrating form $u=\delta$, which  is a cut-off  to regulate the boundary divergence, to $u=u_h=1$.  

As we only consider the expansion of free energy to the order $\epsilon^4$ and $\xi^2$ the above formula of free energy has following properties :

1.  Eq.(4.1) tells us that we need not $\rho_{50}(u)$ nor $\rho_{52}(u)$ term.

2. As $\rho^2\sim\epsilon^2 +{\cal O}(\epsilon^4)$ we see that $At_{42}(u)$ nor $Ax_{41}(u)$ will appear in $ g^{\mu\nu}A_\mu A_\nu\rho^2$. 

3. $At_{42}(u)$ or $Ax_{41}(u)$ could show in $g^{\mu\nu}A_\mu A'_{\nu}$. As it is the value near boundary we only need its leading expansion to $u^2$.

Now, Substituting the found solutions into the on-shell action we find that in the order phase it  has two parts
\be
S_{\rm on-shell}^{\rm order~phase}&=&\Big[-\frac{\epsilon ^2}{4}+\epsilon ^4 \Big(\frac{5}{576}-\frac{\log (2)}{96}\Big)+\xi ^2 \epsilon ^2 \Big(\frac{\log
   (4)}{16}-\frac{1}{16}\Big)\nn\\
&&+\epsilon ^4 \xi ^2 \Big(-\frac{83}{3456}+\frac{5 \pi ^2}{2304}+\frac{5 \log(2)}{288}-\frac{5}{576} \log (2) \log (8)\Big)\Big]\nn\\
&&+\Big[4+\epsilon ^4 \Big(\frac{5 \log (2)}{144}-\frac{331}{13824}\Big)+\frac{\epsilon ^2}{3}\nn\\
&&+\xi ^2\Big(2+ \epsilon^4 \times 0.0078329...+{\log (2)\over 192}+\epsilon ^2 \Big(\frac{2}{9}-\frac{\log (2)}{2}\Big)\Big)\Big]
\ee
where the first bracket part is the on-shell part which needs to do integration while the second bracket part is these obtained without integration.  The on-shell action of disorder phase are the solution with $\rho=0$, which is
\be
S_{\rm on-shell}^{\rm disorder~phase}&=&4+\epsilon ^4 \Big(\frac{7 \log (2)}{288}-\frac{247}{13824}\Big)+\frac{\epsilon ^2}{12}\nn\\
&&+\xi ^2
   \Big(2+ \epsilon ^4\times 0.00814015...+\epsilon ^2 \Big(\frac{23}{144}-\frac{3 \log
   (2)}{8}\Big)\Big)
\ee
Using the on-shell action we can calculate the corresponding free energy.
After subtracting the free energy of disorder phase we get, $\Delta \Omega \equiv \Omega_{\rm order~phase}-\Omega_{\rm disorder~phase}$
\be
\Delta \Omega &=&-{\epsilon ^4\over 384}(1-\xi ^2\times 1.29907...)
\ee
in which $Vol_3$ is the spatial volume of the dual 4D field theory. In the case of $\xi=0$ above result reproduces that in [8]. At the tricritical point $\Delta \Omega =0$ and critical value of flow velocity determined  from above result is
\be
\xi_c&=&0.877372...
\ee
Note that in our method the chemical potential is 
\be
\mu=A_t|_{u=0,\epsilon=0}=2+{\xi_c^2\over 2}=2.38489...
\ee
thus we find that 
\be
{\xi_c\over \mu}=0.367888...
\ee
The exact value obtained in numerical calculation [7] is
\be
 {\xi_c^N\over \mu}=0.274
\ee 
which implies that
\be
\xi_c^N=0.65346
\ee

We see that as the exact value $\xi_c^N$ is not really small our result, which is only expanded to leading order of $\xi$,  could not give a small error therefore. Anyway, it gives a reasonable value because that $0.22645=\Big[\xi_c^N-(\xi_c^N)^2\Big] <\xi_c<\Big[\xi_c^N+(\xi_c^N)^2\Big]=1.08047$.

\section {Conclusion} 
  In this paper we propose a simple trick which enables us to find the exact value of  coefficient in the solution without knowing the exact solution of other function which couples to it.  Our trick leads to the simple formulas (2.45) and (2.46).  We  have tested them in three cases.  Using the crucial trick we have found the higher-order expansion solution in the holographic superfluid with flow and calculated the associated Landau free energy.  We determine the critical value of superfluid velocity at the tricritical point of holographic superfluid and compare it with the numerical value [7]. We see that our analytic method can give a good result. 

  It is expected  that the  method proposed in this paper can be applied to other holographic system and help us to understand the condensed matter physics from  AdS/CFT correspondence.
\\
\\
\begin{center} {\bf REFERENCES}\end{center}
\begin{enumerate}
\item  J. M. Maldacena, ``'Large N limit of superconformal field theories and supergravity," Adv. Theor. Math. Phys. 2 (1998) 231  [Int. J. Theor. Phys. 38 (1999) 1113 ] [arXiv:hep-th/9711200].
\item S. S. Gubser, I. R. Klebanov and A. M. Polyakov, ``Gauge theory theory correlators from non-critical string theory," Phys. Lett. B428 (1998) 105  [arXiv:hep-th/9802109].
\item  E. Witten, ``Anti-de Sitter space and holography," Adv. Theor. Math. Phys. 2 (1998) 253  [arXiv:hepth/9802150]. 
\item C. P. Herzog, P. Kovtun, S. Sachdev and D. T. Son, ``Quantum critical transport, duality, and M theory," Phys. Rev. D75 (2007) 085020  [arXiv:hep-th/0701036].
\item S. S. Gubser, ``Breaking an Abelian gauge symmetry near a black hole horizon," Phys. Rev. D78 (2008) 065034  [arXiv:0801.2977 [hep-th]]; S. S. Gubser, ``Colorful horizons with charge in anti-de Sitter space," Phys. Rev. Lett. 101 (2008) 191601  [arXiv:0803.3483 [hep-th]].
\item S. A. Hartnoll, C. P. Herzog and G. T. Horowitz, ``Holographic Superconductors," JHEP 0812 (2008) 015  [arXiv:0810.1563 [hep-th]].
\item C. P. Herzog, P. K. Kovtun and D. T. Son, ``Holographic model of superfluidity," Phys. Rev. D79 (2009) 066002 [arXiv:0809.4870 [hep-th]].
\item C. P. Herzog, ``An Analytic Holographic Superconductor," Phys.Rev.D81 (2010) 126009 [arXiv:1003.3278 [hep-th]].
\item C. P. Herzog and S. S. Pufu, ``The Second Sound of SU(2)," JHEP 0904 (2009) 126.  [arXiv:0902.0409 [hep-th]].
\item Wung-Hong Huang, ``Analytic Study of First-Order Phase Transition in Holographic Superconductor and Superfluid," Int. J. Mod. Phys. A 28 (2013) [arXiv:1307.5614 [hep-th]]. 
\item R. Gregory, S. Kanno and J. Soda, ``Holographic Superconductors with Higher Curva ture Corrections," JHEP 0910 (2009) 010  [arXiv:0907.3203 [hep-th]].
\item L. D. Landau and E. M. Lifshitz, ``Statistical Physics,"  2005, Pergamon Press. 
\end{enumerate}
\end{document}